\documentclass[a4paper, oneside, twocolumn, notitlepage, 10pt]{extarticle_ecoc}
\pdfoutput=1
\usepackage{ecoc}
\usepackage{fst}

\usepackage{subcaption}

\begin{document}
%\selectlanguage{english}    

\title{Comparison of PAM-6 Modulations for Short-Reach Fiber-Optic Links with Intensity Modulation and Direct Detection}

\author{
    Tobias Prinz\textsuperscript{(1),$\dagger$}, Thomas Wiegart\textsuperscript{(1)}, Daniel Plabst\textsuperscript{(1)},
    Talha Rahman\textsuperscript{(2)}, Md Sabbir-Bin Hossain\textsuperscript{(2)},\\  Neboj\v{s}a Stojanovi\'{c}\textsuperscript{(2)}, Stefano Calabr\`{o}\textsuperscript{(2)},
    Norbert Hanik\textsuperscript{(1)}, Gerhard Kramer\textsuperscript{(1)}
}

\maketitle

\begin{strip}
 \begin{author_descr}

    \centering
   \textsuperscript{(1)} Institute for Communications Engineering, Technical University of Munich, 80333, Munich, Germany

   \textsuperscript{(2)} Huawei Technologies Duesseldorf GmbH, Munich Research Center, 80992, Munich, Germany
   
\textsuperscript{$\dagger$}Email: \textcolor{blue}{\uline{tobias.prinz@tum.de}}

 \end{author_descr}
\end{strip}

\setstretch{1.1}

\begin{strip}
  \begin{ecoc_abstract}
    PAM-6 transmission is considered for short-reach fiber-optic links with intensity modulation and direct detection. Experiments show that probabilistically-shaped PAM-6 and a framed-cross QAM-32 constellation outperform conventional cross \mbox{QAM-32} under a peak power constraint. 
  \end{ecoc_abstract}
\end{strip}

\section{Introduction}
Short-reach fiber-optic transceivers often use intensity modulation (IM) with direct detection (DD)~\cite{zhong18}. The connecting links are usually operated without optical amplifiers, resulting in peak power constrained systems~\cite{wiegart20}. For these systems, it was shown in~\cite{prinz21} that pulse-amplitude modulation with six levels (PAM-6) outperforms PAM-4 and PAM-8 at rates around 2~bits per channel use~(bpcu).

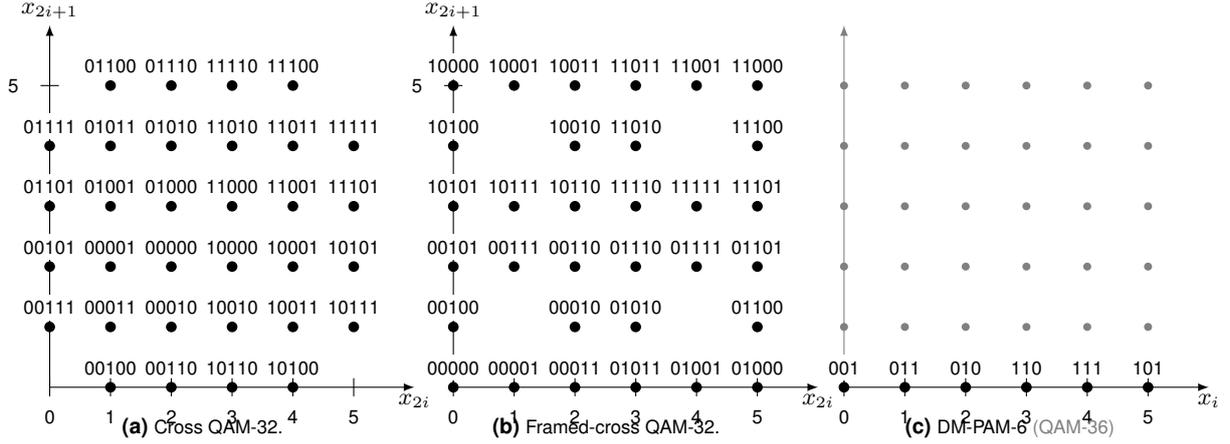
\begin{figure*}[t] %
        \begin{subfigure}[b]{0.33\textwidth}
        \centering 
        \tikzstyle{point}=[fill,shape=circle,minimum size=4pt,inner sep=0pt]
\begin{tikzpicture}[scale=0.8]
\scriptsize
	\draw[-latex] (0,0) -- (6,0) node[right,below] {\small $x_{2i}$};
	\draw[-latex] (0,0) -- (0,6) node[above] {\small $x_{2i+1}$};

	\draw (0,0.15) -- (0,-0.15) node[below,yshift=-0.1cm] {0};
	\draw (1,0.15) -- (1,-0.15) node[below,yshift=-0.1cm] {1};
	\draw (2,0.15) -- (2,-0.15) node[below,yshift=-0.1cm] {2};
	\draw (3,0.15) -- (3,-0.15) node[below,yshift=-0.1cm] {3};
	\draw (4,0.15) -- (4,-0.15) node[below,yshift=-0.1cm] {4};
	\draw (5,0.15) -- (5,-0.15) node[below,yshift=-0.1cm] {5};
	\draw (0.15,5) -- (-0.15,5) node[left,xshift=-0.2cm] {5};

	\node[point,label={[label distance=0.1]90:{00100}}] at (1,0) {};
	\node[point,label={[label distance=0.1]90:{00110}}] at (2,0) {};
	\node[point,label={[label distance=0.1]90:{10110}}] at (3,0) {};
	\node[point,label={[label distance=0.1]90:{10100}}] at (4,0) {};
	
	\node[point,label={[fill=white,label distance=0.1]90:{00111}}] at (0,1) {};
	\node[point,label={[label distance=0.1]90:{00011}}] at (1,1) {};
	\node[point,label={[label distance=0.1]90:{00010}}] at (2,1) {};
	\node[point,label={[label distance=0.1]90:{10010}}] at (3,1) {};
	\node[point,label={[label distance=0.1]90:{10011}}] at (4,1) {};
	\node[point,label={[label distance=0.1]90:{10111}}] at (5,1) {};
	
	\node[point,label={[fill=white,label distance=0.1]90:{00101}}] at (0,2) {};
	\node[point,label={[label distance=0.1]90:{00001}}] at (1,2) {};
	\node[point,label={[label distance=0.1]90:{00000}}] at (2,2) {};
	\node[point,label={[label distance=0.1]90:{10000}}] at (3,2) {};
	\node[point,label={[label distance=0.1]90:{10001}}] at (4,2) {};
	\node[point,label={[label distance=0.1]90:{10101}}] at (5,2) {};
	
	\node[point,label={[fill=white,label distance=0.1]90:{01101}}] at (0,3) {};
	\node[point,label={[label distance=0.1]90:{01001}}] at (1,3) {};
	\node[point,label={[label distance=0.1]90:{01000}}] at (2,3) {};
	\node[point,label={[label distance=0.1]90:{11000}}] at (3,3) {};
	\node[point,label={[label distance=0.1]90:{11001}}] at (4,3) {};
	\node[point,label={[label distance=0.1]90:{11101}}] at (5,3) {};
	
	\node[point,label={[fill=white,label distance=0.1]90:{01111}}] at (0,4) {};
	\node[point,label={[label distance=0.1]90:{01011}}] at (1,4) {};
	\node[point,label={[label distance=0.1]90:{01010}}] at (2,4) {};
	\node[point,label={[label distance=0.1]90:{11010}}] at (3,4) {};
	\node[point,label={[label distance=0.1]90:{11011}}] at (4,4) {};
	\node[point,label={[label distance=0.1]90:{11111}}] at (5,4) {};
	
	\node[point,label={[label distance=0.1]90:{01100}}] at (1,5) {};
	\node[point,label={[label distance=0.1]90:{01110}}] at (2,5) {};
	\node[point,label={[label distance=0.1]90:{11110}}] at (3,5) {};
	\node[point,label={[label distance=0.1]90:{11100}}] at (4,5) {};

\end{tikzpicture}
        \vspace{-0.8cm}
        \caption{Cross QAM-32.}
        \label{fig:cross}
        \end{subfigure}\hfill%
        %-----------
        \begin{subfigure}[b]{0.33\textwidth}
        \centering 
        \tikzstyle{point}=[fill,shape=circle,minimum size=4pt,inner sep=0pt]
\begin{tikzpicture}[scale=0.8]
\scriptsize
	\draw[-latex] (0,0) -- (6,0) node[right,below] {\small $x_{2i}$};
	\draw[-latex] (0,0) -- (0,6) node[above] {\small $x_{2i+1}$};
	\draw (0,0.15) -- (0,-0.15) node[below,yshift=-0.1cm] {0};
	\draw (1,0.15) -- (1,-0.15) node[below,yshift=-0.1cm] {1};
	\draw (2,0.15) -- (2,-0.15) node[below,yshift=-0.1cm] {2};
	\draw (3,0.15) -- (3,-0.15) node[below,yshift=-0.1cm] {3};
	\draw (4,0.15) -- (4,-0.15) node[below,yshift=-0.1cm] {4};
	\draw (5,0.15) -- (5,-0.15) node[below,yshift=-0.1cm] {5};
	\draw (0.15,5) -- (-0.15,5) node[left,xshift=-0.2cm] {5};
	\node[point,label={[fill=white,label distance=0.1]90:{00000}}] at (0,0) {};
	\node[point,label={[label distance=0.1]90:{00001}}] at (1,0) {};
	\node[point,label={[label distance=0.1]90:{00011}}] at (2,0) {};
	\node[point,label={[label distance=0.1]90:{01011}}] at (3,0) {};
	\node[point,label={[label distance=0.1]90:{01001}}] at (4,0) {};
	\node[point,label={[label distance=0.1]90:{01000}}] at (5,0) {};
	
	\node[point,label={[fill=white,label distance=0.1]90:{00100}}] at (0,1) {};
	\node[point,label={[label distance=0.1]90:{00010}}] at (2,1) {};
	\node[point,label={[label distance=0.1]90:{01010}}] at (3,1) {};
	\node[point,label={[label distance=0.1]90:{01100}}] at (5,1) {};
	
	\node[point,label={[fill=white,label distance=0.1]90:{00101}}] at (0,2) {};
	\node[point,label={[label distance=0.1]90:{00111}}] at (1,2) {};
	\node[point,label={[label distance=0.1]90:{00110}}] at (2,2) {};
	\node[point,label={[label distance=0.1]90:{01110}}] at (3,2) {};
	\node[point,label={[label distance=0.1]90:{01111}}] at (4,2) {};
	\node[point,label={[label distance=0.1]90:{01101}}] at (5,2) {};
	
	\node[point,label={[fill=white,label distance=0.1]90:{10101}}] at (0,3) {};
	\node[point,label={[label distance=0.1]90:{10111}}] at (1,3) {};
	\node[point,label={[label distance=0.1]90:{10110}}] at (2,3) {};
	\node[point,label={[label distance=0.1]90:{11110}}] at (3,3) {};
	\node[point,label={[label distance=0.1]90:{11111}}] at (4,3) {};
	\node[point,label={[label distance=0.1]90:{11101}}] at (5,3) {};
	
	\node[point,label={[fill=white,label distance=0.1]90:{10100}}] at (0,4) {};
	\node[point,label={[label distance=0.1]90:{10010}}] at (2,4) {};
	\node[point,label={[label distance=0.1]90:{11010}}] at (3,4) {};
	\node[point,label={[label distance=0.1]90:{11100}}] at (5,4) {};
	
	\node[point,label={[fill=white,label distance=0.1]90:{10000}}] at (0,5) {};
	\node[point,label={[label distance=0.1]90:{10001}}] at (1,5) {};
	\node[point,label={[label distance=0.1]90:{10011}}] at (2,5) {};
	\node[point,label={[label distance=0.1]90:{11011}}] at (3,5) {};
	\node[point,label={[label distance=0.1]90:{11001}}] at (4,5) {};
	\node[point,label={[label distance=0.1]90:{11000}}] at (5,5) {};
\end{tikzpicture}
        \vspace{-0.8cm}
        \caption{Framed-cross QAM-32.}
        \label{fig:framed_cross}
        \end{subfigure}\hfill%
        %-----------
        \begin{subfigure}[b]{0.33\textwidth}
        \centering 
        \tikzstyle{point}=[fill,shape=circle,minimum size=4pt,inner sep=0pt]
\begin{tikzpicture}[scale=0.8]
\scriptsize
	\draw[-latex] (0,0) -- (6,0) node[right,below] {\small $x_{i}$};
	\draw[-latex,gray] (0,0) -- (0,6) node[above,white] {\small $x_{2i+1}$};
	\draw (0,0.15) -- (0,-0.15) node[below,yshift=-0.1cm] {0};
	\draw (1,0.15) -- (1,-0.15) node[below,yshift=-0.1cm] {1};
	\draw (2,0.15) -- (2,-0.15) node[below,yshift=-0.1cm] {2};
	\draw (3,0.15) -- (3,-0.15) node[below,yshift=-0.1cm] {3};
	\draw (4,0.15) -- (4,-0.15) node[below,yshift=-0.1cm] {4};
	\draw (5,0.15) -- (5,-0.15) node[below,yshift=-0.1cm] {5};

	\node[point,label={[fill=white,label distance=0.1]90:{001}}] at (0,0) {};
	\node[point,label={[label distance=0.1]90:{011}}] at (1,0) {};
	\node[point,label={[label distance=0.1]90:{010}}] at (2,0) {};
	\node[point,label={[label distance=0.1]90:{110}}] at (3,0) {};
	\node[point,label={[label distance=0.1]90:{111}}] at (4,0) {};
	\node[point,label={[label distance=0.1]90:{101}}] at (5,0) {};

	\node[point,gray,minimum size=3pt] at (0,1) {};
	\node[point,gray,minimum size=3pt] at (1,1) {};
	\node[point,gray,minimum size=3pt] at (2,1) {};
	\node[point,gray,minimum size=3pt] at (3,1) {};
	\node[point,gray,minimum size=3pt] at (4,1) {};
	\node[point,gray,minimum size=3pt] at (5,1) {};
	
	\node[point,gray,minimum size=3pt] at (0,2) {};
	\node[point,gray,minimum size=3pt] at (1,2) {};
	\node[point,gray,minimum size=3pt] at (2,2) {};
	\node[point,gray,minimum size=3pt] at (3,2) {};
	\node[point,gray,minimum size=3pt] at (4,2) {};
	\node[point,gray,minimum size=3pt] at (5,2) {};
	
	\node[point,gray,minimum size=3pt] at (0,3) {};
	\node[point,gray,minimum size=3pt] at (1,3) {};
	\node[point,gray,minimum size=3pt] at (2,3) {};
	\node[point,gray,minimum size=3pt] at (3,3) {};
	\node[point,gray,minimum size=3pt] at (4,3) {};
	\node[point,gray,minimum size=3pt] at (5,3) {};
	
	\node[point,gray,minimum size=3pt] at (0,4) {};
	\node[point,gray,minimum size=3pt] at (1,4) {};
	\node[point,gray,minimum size=3pt] at (2,4) {};
	\node[point,gray,minimum size=3pt] at (3,4) {};
	\node[point,gray,minimum size=3pt] at (4,4) {};
	\node[point,gray,minimum size=3pt] at (5,4) {};
	
	\node[point,gray,minimum size=3pt] at (0,5) {};
	\node[point,gray,minimum size=3pt] at (1,5) {};
	\node[point,gray,minimum size=3pt] at (2,5) {};
	\node[point,gray,minimum size=3pt] at (3,5) {};
	\node[point,gray,minimum size=3pt] at (4,5) {};
	\node[point,gray,minimum size=3pt] at (5,5) {};

\end{tikzpicture}
        \vspace{-0.8cm}
        \caption{\mbox{DM-PAM-6} {\color{gray}(QAM-36)}}
        \label{fig:pam6dm}
        \end{subfigure}
        %-----------
    \vspace{0.5cm}
    \caption{Three constellations that allow transmission of PAM-6 alphabets with time index $i\in\mathbb{N}$.}
    \vspace{0.4cm}
    \label{fig:2x3_plot_L=0km}
\end{figure*}

Cross-shaped QAM-32 (\mbox{C-QAM-32}) constellations (cf.~Fig.~\ref{fig:cross}) can be used to generate PAM-6 symbols~\cite{ghiasi12,chorchos19}. Using \mbox{C-QAM-32} with a peak power constraint is sub-optimal under additive white Gaussian noise (AWGN)~\cite{prinz21}. Other options for generating PAM-6 are a framed-cross QAM-32 (\mbox{FC-QAM-32}) constellation\cite{prinz21} (cf. Fig.~\ref{fig:framed_cross}) and PAM-6 (QAM-36) generated with probabilistic amplitude shaping (PAS)\cite{boecherer15} (cf. Fig.~\ref{fig:pam6dm}).

In~\cite{prinz21} it was shown that \mbox{FC-QAM-32} and PAM-6 via PAS gain \SI{0.4}{dB} SNR over \mbox{C-QAM-32} under symbol-metric decoding and \SI{0.8}{dB} SNR under bit-metric decoding, when transmitting at a rate of \SI{2}{bpcu} over an AWGN channel.

In this paper, we conduct experiments at a symbol rate of \SI{60}{GBaud} over \SI{1}{km} of standard single-mode fiber (SSMF) and optical back-to-back (B2B)  experiments at a symbol rate of \SI{95.6}{GBaud}. We show that the gains from~\cite{prinz21} also apply to practical scenarios.
The measured rate gains are not quite as large as for the AWGN case\cite{prinz21} but they nevertheless motivate using \mbox{FC-QAM-32} rather than \mbox{C-QAM-32}.

\begin{figure*}[!t]
    \centering
    \includegraphics[width=\textwidth]{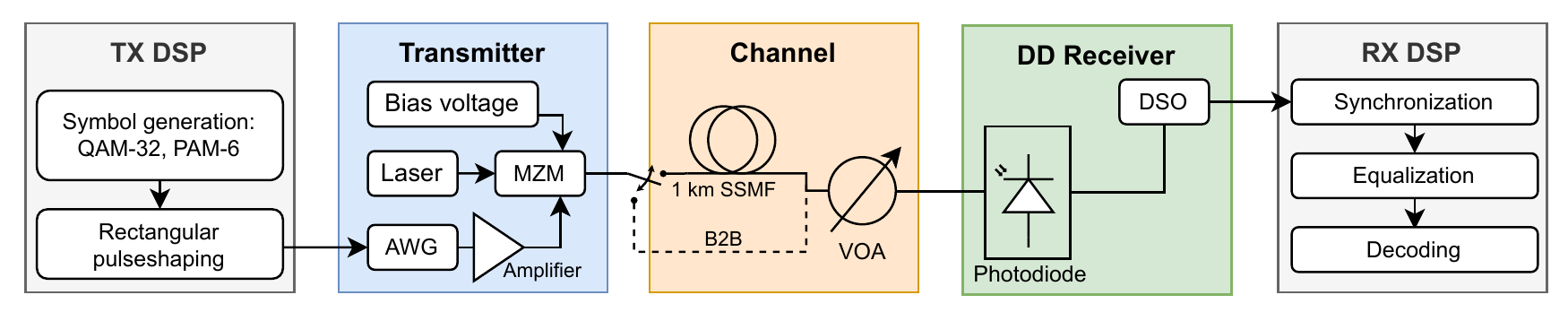}
    \caption{Experimental setup.}
    \label{fig:experiment}
\end{figure*}

\section{PAM-6 Transmission}
One cannot map bits directly to symbols from an input alphabet with cardinality six, since six is not an integer power of two. In the following, we present two ways to generate a sequence from a 6-ary input alphabet. We define the input alphabet as $\mathcal{X}=\{ 0,1,2,3,4,5 \}$.

\section{PAM-6 via QAM-32}
Consider two-dimensional QAM-32 constellations for which 5 bits are directly mapped to a symbol. The ``real'' and ``imaginary'' parts of the symbols can be transmitted consecutively as PAM-6 symbols. \mbox{C-QAM-32} is proposed for PAM-6 transmission in\cite{ghiasi12,chorchos19}, but~\cite{prinz21} showed that \mbox{C-QAM-32} is sub-optimal under peak power constrained AWGN channels. Furthermore, there exists no Gray labeling for \mbox{C-QAM-32}, which results in a performance loss under bit-metric decoding.

The proposed \mbox{FC-QAM-32} constellation~\cite{prinz21} is shown in Fig.~\ref{fig:framed_cross} with according Gray labeling. By using the corner levels of the constellation, \mbox{FC-QAM-32} does not increase the peak power, but provides a larger average Euclidean distance of the symbols at the receiver. This significantly increases the performance in AWGN channels. 
The maximum achievable information rate of both QAM-32 schemes is $\log_2(32)/2=\SI{2.5}{bpcu}$. 

\section{PAM-6 via Distribution Matcher}
Probabilistic amplitude shaping (PAS)~\cite{boecherer15} for \mbox{PAM-6} with a distribution matcher (DM) is studied in~\cite{prinz21}. The same idea can be used to construct flexible QAM constellations\cite{buchali2017spectrally} or to label QAM-32 with 6 bits~\cite{boecherer_2014_labeling}. The scheme is denoted as \mbox{DM-PAM-6} and operates as follows. 

Consider a source that outputs $k+\gamma n$ uniformly distributed data bits 
\begin{equation}
    \boldsymbol{d} = ( d_1^k, d_{k+1}^{k+\gamma n} ) 
\end{equation}
consisting of a length $k$ sequence $d_1^k$, and a length $\gamma n$ sequence $d_{k+1}^{k+\gamma n}$, where $n$ is the number of PAM-6 symbols in a forward error correction (FEC) frame. A DM maps the source sequence $d_1^k$ to the sequence $a_1^n$, where $\forall i, a_i \in \mathcal{A}$ and ternary alphabet $\mathcal{A}=\{0,1,2\}$. We choose a uniform distribution for the distribution matcher, i.e., $P_A(a) = 1/3$ for all $a \in \mathcal{A}$.
The elements in $\mathcal{A}$ are labeled as bit-pairs from the set $\{01,11,10\}$  which correspond to the second and third bit in the labeling in Fig.~\ref{fig:pam6dm}. Labeling every symbol of the sequence $a_1^n = (a_1,\ldots,a_n)$ results in a length $2n$ bit sequence $\boldsymbol{b} = (\boldsymbol{b}_1,\ldots,\boldsymbol{b}_n)$, where $\boldsymbol{b}_i$ is the length two bit label of the symbol $a_i$. We concatenate $\boldsymbol{b}$ with the second part $d_{k+1}^{k+\gamma n}$ of the source bits and obtain $\boldsymbol{u} = (\boldsymbol{b},d_{k+1}^{k+\gamma n})$ 
% \begin{equation}
%     \boldsymbol{u} = \left[\boldsymbol{b},u_{k+1}^{k+\gamma n}\right]
% \end{equation}
of length $(2+\gamma)n$ bits. We feed $\boldsymbol{u}$ into a systematic FEC encoder that generates $(1-\gamma)n$ parity bits $\boldsymbol{p}$. The code rate of the FEC code is
\begin{equation}
R_\mathrm{FEC} = \frac{(2+\gamma)n}{3n} = \frac{2+\gamma}{3}.    
\end{equation}
We assume that the parity bits in $\boldsymbol{p}$ are uniformly distributed~\cite{boecherer15}. We concatenate the $(1-\gamma)n$ parity bits $\boldsymbol{p}$ with the source sequence $d_{k+1}^{k+\gamma n}$ and obtain the length $n$ sequence $\boldsymbol{s} = (s_1,\ldots,s_n)$ with uniformly distributed entries. The bit entries $s_i \in \boldsymbol{s}$ are used as the first bit (MSB) in the labeling in Fig.~\ref{fig:pam6dm}. Finally, we concatenate the $s_i$ with the label $\boldsymbol{b}_i$ to obtain the complete 3-bit label of the $i^\text{th}$ transmit symbol and map the label to the corresponding symbol from the input alphabet $\mathcal{X}$.

Similar to the QAM-32 schemes, the presented \mbox{DM-PAM-6} scheme can also be interpreted as QAM-36 in two-dimensions as illustrated in Fig.~\ref{fig:pam6dm}. Note that \mbox{DM-PAM-6} also permits non-uniform input distributions (see \cite{prinz21}) which allows further optimizations.

The transmission rate of \mbox{DM-PAM-6} is
\begin{align}
  R = \frac{k}{n} + \gamma \,\si{bpcu}. \label{eq:rate}
\end{align}
For large block lengths $n$ and a suitable DM, the rate $k/n$ approaches the entropy $\mathbb{H}(P_A)$ of the density $P_A$. For uniform $P_A$, $\mathbb{H}(P_A) \approx 1.585\,$bits. The \mbox{DM-PAM-6} rate $R$ in~\eqref{eq:rate} ranges from $R \approx \SI{1.585}{bpcu}$ to \mbox{$R=\mathbb{H}(P_A)+1 \approx \SI{2.585}{bpcu}$} for $\gamma =0$ and $\gamma=1$, respectively. For finite block lengths $n$, $R$ decreases due to the rate loss of the DM.

\section{Experimental Setup}
\begin{figure*}
{
    \begin{subfigure}[b]{0.5\textwidth}
        \centering
        \begin{tikzpicture} 
\pgfplotsset{
  set layers,
}
\pgfplotsset{ mark layer=axis tick labels}

\footnotesize
\begin{axis}[set layers,
xlabel={Attenuation [\si{dB}]},
width=\textwidth, 
height=6.3cm,
ylabel={Achievable information rate (bpcu)},
grid=both,
legend cell align={left},
 legend style ={at={(1,0)},anchor=south east, text opacity=1},
 xmin=2,
 xmax=8,
 ymin=1.7,
 ymax=2.6,
 x dir=reverse
]

\addplot[name path global=cross,styleI] table[x=attenuation,y=rate] {results/BCH_PAM6Cross_60GBaud_complex_bcjr_n=1000.txt};
\addlegendentry{\scriptsize C-QAM-32};

\addplot[name path global=fcross,styleII] table[x=attenuation,y=rate] {results/BCH_PAM6FramedCross_60GBaud_complex_bcjr_n=1000.txt};
\addlegendentry{\scriptsize FC-QAM-32};

\addplot[name path global=pam6dm,styleIII] table[x=attenuation,y=rate] {results/BCH_PAM6DM_60GBaud_complex_bcjr_n=1000.txt};
\addlegendentry{\scriptsize DM-PAM-6};

\addplot[name path global=ldpc_cross,styleI,LDPC] table[x=attenuation,y=rate] {results/LDPC_PAM6Cross_60GBaud_complex_bcjr_n=1000.txt};

\addplot[name path global=ldpc_fcross,styleII,LDPC] table[x=attenuation,y=rate] {results/LDPC_PAM6FramedCross_60GBaud_complex_bcjr_n=1000.txt};

\addplot[name path global=ldpc_pam6dm,styleIII,LDPC] table[x=attenuation,y=rate] {results/LDPC_PAM6DM_60GBaud_complex_bcjr_n=1000.txt};

\path[name path global=line] (axis cs:\pgfkeysvalueof{/pgfplots/xmin},2) -- (axis cs: \pgfkeysvalueof{/pgfplots/xmax},2);
\path[name intersections={of=line and pam6dm, name=p1}, name intersections={of=line and cross, name=p2}];

\draw[arr,shadowed] (axis cs:5.55,2) -- (axis cs:5.27,2) node[font=\footnotesize,right] {\SI{0.3}{dB}};

\draw[arr,shadowed] (axis cs:7.25,1.985) -- (axis cs:6.76,1.985) node[font=\footnotesize,right,yshift=-2pt] {\SI{0.5}{dB}};

\draw[arr,shadowed] (axis cs:6.98,2.02) -- (axis cs:6.62,2.02) node[font=\footnotesize,right,yshift=+2pt] {\SI{0.4}{dB}};

\end{axis}

\end{tikzpicture}
        \caption{Symbol rate: \SI{60}{GBaud}, \SI{1}{km}.}
        \label{fig:bch_60}
    \end{subfigure}
    \begin{subfigure}[b]{0.5\textwidth}
        \centering
        \begin{tikzpicture} 
\pgfplotsset{
  set layers,
}
\pgfplotsset{ mark layer=axis tick labels}

\footnotesize
\begin{axis}[
xlabel={Attenuation [\si{dB}]},
width=\textwidth, 
height=6.3cm,
ylabel={Achievable information rate (bpcu)},
grid=both,
legend cell align={left},
 legend style ={at={(1,0)},anchor=south east,  text opacity=1},
 xmin=2,
 xmax=8,
 ymin=1.7,
 ymax=2.6,
 x dir=reverse
]

\addplot[name path global=cross,styleI] table[x=attenuation,y=rate] {results/BCH_PAM6Cross_95GBaud_complex_bcjr_n=1000.txt};
\addlegendentry{\scriptsize C-QAM-32};

\addplot[name path global=fcross,styleII] table[x=attenuation,y=rate] {results/BCH_PAM6FramedCross_95GBaud_complex_bcjr_n=1000.txt};
\addlegendentry{\scriptsize FC-QAM-32};

\addplot[name path global=pam6dm,styleIII] table[x=attenuation,y=rate] {results/BCH_PAM6DM_95GBaud_complex_bcjr_n=1000.txt};
\addlegendentry{\scriptsize DM-PAM-6};

\addplot[name path global=ldpc_cross,LDPC,styleI] table[x=attenuation,y=rate] {results/LDPC_PAM6Cross_95GBaud_complex_bcjr_n=1000.txt};

\addplot[name path global=ldpc_fcross,LDPC,styleII] table[x=attenuation,y=rate] {results/LDPC_PAM6FramedCross_95GBaud_complex_bcjr_n=1000.txt};

\addplot[name path global=ldpc_pam6dm,LDPC,styleIII] table[x=attenuation,y=rate] {results/LDPC_PAM6DM_95GBaud_complex_bcjr_n=1000.txt};

\draw[arr,shadowed] (axis cs:4.92,2.0) -- (axis cs:4.68,2.0) node[font=\footnotesize,right] {\SI{0.2}{dB}};

\draw[arr,shadowed] (axis cs:6.48,2.0) -- (axis cs:6.15,2.0) node[font=\footnotesize,right] {\SI{0.35}{dB}};

\end{axis}

\end{tikzpicture}
        \caption{Symbol rate: \SI{95.6}{GBaud}, optical B2B.}
        \label{fig:bch_95}
    \end{subfigure}
    }
    \vspace{0cm}
    \caption{Achievable rates of the three schemes using BCH (solid lines) and LDPC codes (dotted lines) with length $n=1000$ real-valued channel uses evaluated at an FER of $10^{-3}$.}
    \label{fig:bch}
\end{figure*}
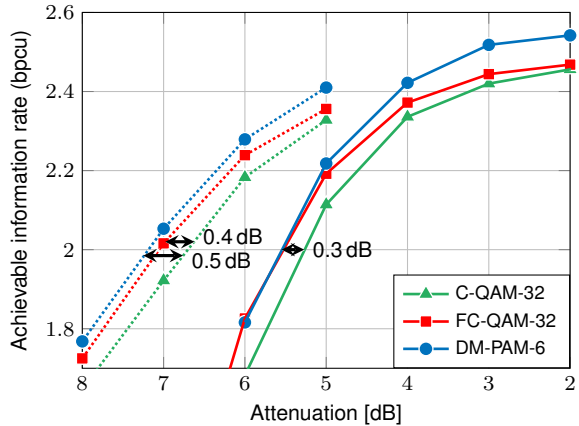
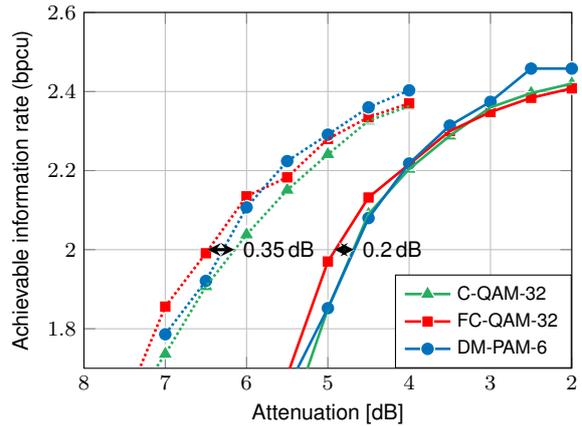
The experimental setup is depicted in Fig.~\ref{fig:experiment}. The transmitter generates constellation symbols and performs rectangular pulse shaping. 
The arbitrary waveform generator (AWG) operates at sampling rates $\SI{95.6}{GSamples/s}$ and $\SI{120}{GSamples/s}$. The first sampling rate was used for $\SI{95.6}{GBaud}$ transmission with 1 sample per symbol (SPS) and the latter for $\SI{60}{GBaud}$ transmission with 2 SPS.
The AWG signal is electrically amplified and drives a Mach-Zehnder modulator (MZM) that modulates a $\SI{1550}{nm}$ external cavity laser. The MZM is operated at the quadrature point. We consider either transmission over $\SI{1}{km}$ of SSMF or an optical back-to-back (B2B) configuration. Both setups are connected to a variable optical attenuator (VOA). 

After the VOA, a photodiode converts the optical intensity to an electrical current that is sampled by a digital storage oscilloscope (DSO) operating at $\SI{256}{GSamples/s}$ and the samples are stored for offline processing. We perform synchronization and resample to $\SI{2}{SPS}$ before equalization with a Volterra equalizer, using 41 first-order, 7 second-order and 5 third-order taps. The equalizer is trained based on an MSE objective. The equalized sequence is downsampled to the symbol rate to retrieve the symbols. For the setup with and without fiber, measurements are taken for different attenuation factors of the VOA. For \SI{95.6}{GBaud} symbol rate transmission, we use an additional 3-tap noise whitening filter\cite{wettlin20} and compute the symbol-wise a posteriori probabilities through the BCJR algorithm\cite{bcjr}.

\section{Experimental Results with FEC}
We compare the described methods in combination with forward error correction (FEC). We consider decoding based on hard-decisions and decoding based on soft-information. Since the transmitted sequences from the experiments are usually not codewords of a FEC code, we apply the method of channel adapters and scrambling sequences~\cite{boecherer19}. 
 
We consider blocks of $n=1000$ symbols. This results in a binary codelength of \SI{2500}{bits} and \SI{3000}{bits} for QAM-32 and \mbox{DM-PAM-6}, respectively. We use shortened BCH codes with a mother code length of \SI{4095}{bits} for hard-decision decoding and low-density parity-check (LDPC) codes from the 5G standard for soft-decoding~\cite{ldpc5g}. The transmission rate is adjusted by changing the code dimension. For \mbox{DM-PAM-6}, we use a constant composition distribution matcher (CCDM)~\cite{schulte15}. 

Fig.~\ref{fig:bch} shows achievable information rates for transmissions at \SI{60}{GBaud} symbol rate over \SI{1}{km} SSMF, and at \SI{95.6}{GBaud} symbol rate in optical B2B setup. For each attenuation factor, we adjust the transmission rate such that the frame error rate (FER) is $10^{-3}$. The results show that \mbox{FC-QAM-32} gains between \SI{0.2}{dB} and \SI{0.4}{dB} over conventional \mbox{C-QAM-32} for each scenario, when operating at a rate of \SI{2}{bpcu}. The largest gains are for bit-metric soft-decoding with LDPC codes. Especially the \mbox{FC-QAM-32} gains are noteworthy, since implementation complexity is the same as for \mbox{C-QAM-32}. \mbox{DM-PAM-6} gains up to \SI{0.5}{dB} over conventional \mbox{C-QAM-32}. For \SI{95.6}{GBaud}, \mbox{DM-PAM-6} shows smaller gains, especially for high attenuation.

\section{Conclusions}
We compared three methods for generating \mbox{PAM-6} in IM/DD link. Experiments with \SI{1}{km} SSMF and optical B2B transmission show that \mbox{FC-QAM-32} gains up to \SI{0.4}{dB} in power over \mbox{C-QAM-32}. The framed-cross constellation allows Gray labeling and adds no complexity compared to \mbox{C-QAM-32}. PAM-6 with a DM gains up to \SI{0.5}{dB} over conventional PAM-6, but suffers for high symbol rates where bandwidth limitations from AWG and photodiode increase. Moreover, the \mbox{DM-PAM-6} scheme has a higher complexity due to the DM.

\clearpage
\printbibliography
\end{document}